\documentstyle[twocolumn,prb,aps,epsf]{revtex}

\begin{document}
\draft

\twocolumn[\hsize\textwidth\columnwidth\hsize\csname @twocolumnfalse\endcsname

\title{
Coherence Resonance and Noise-Induced Synchronization in Globally
Coupled Hodgkin-Huxley Neurons
}
\author{Yuqing Wang, David T. W. Chik, and Z. D. Wang}

\address{
Department of Physics, The University of Hong Kong, \\
Pokfulam Road, Hong Kong, P.R. China \\
}

\maketitle

\begin{abstract}
The coherence resonance (CR) of globally coupled Hodgkin-Huxley neurons
is studied. When the neurons are set in the
subthreshold regime near the firing threshold, the additive noise induces
limit cycles.  The coherence of the system is optimized by the noise.
A bell-shaped curve is found for the peak height of power spectra of the
spike train, being significantly different from a monotonic behavior for the
single neuron.
The coupling of the network can enhance CR in two different ways.
In particular, when the coupling is strong enough, the synchronization
of the system is induced and optimized by the noise.
This synchronization leads to a high and wide
plateau in the local measure of coherence curve. The local-noise-induced limit cycle
can evolve to a refined spatiotemporal order through the dynamical
optimization among the autonomous oscillation of an individual neuron,
the coupling of the network, and the local noise.
\end{abstract}


\vskip2pc
]

\newpage 
The phenomenon of stochastic resonance (SR) has been intensively studied
for the last decade\cite{R1}. The response of a noisy nonlinear
system to a deterministic signal can be optimized by 
noise. Recently, it has been shown that, in the absence of
a deterministic signal, the noisy nonlinear system exhibits SR-like
behavior \cite{R2,R3,R4,R5,R6,R7,R8}. This phenomenon, which is
referred to as coherence resonance (CR) or autonomous SR, was
first discussed in a simple autonomous system in the vicinity
of the saddle-node bifurcation \cite{R2,R3}. The nonuniform
noise-induced limit cycle leads to a peak  at a definite
frequency in the power spectrum. The signal-to-noise ratio (SNR)
increases first to a maximum and then decreases when
the intensity of noise increases, showing the
optimization of the coherent limit cycle to the noise.
The frequency was observed to shift to a  higher value by
increasing the noise intensity. The CR has also been found
in excitable systems, {\sl e. g.,} the Fitz Hugh-Nagumo
model \cite{R4}, the Hodgkin-Huxley (HH) model \cite{R5}, the
Plant model and the Hindermarsh-Rose model\cite{R6}.
Moreover, an experimental evidence of CR was reported
very recently\cite{R8}.

Synchronization in nonlinear stochastic systems has 
also attracted growing interests in recent 
years\cite{R8a,R8b,R8c,R9,R10,R11,R12}.
SR and noise-induced global synchronization
have been studied.  Regardless of whether the system is locally or globally coupled, 
the coupling can enhance the signal
transduction and the SNR of the local unit. The coupling strength
can be considered to be another tuning parameter of SR. Meanwhile,
the noise-induced global synchronization, which coincides with the
optimized local performance of the single element in the network,
is observed.  Moreover,  in the study of the coupled stochastic limit cycle,
Kurrer and Schulten \cite{R12a} have studied analytically a model of globally coupled
stochastic neurons and found noise-enhanced synchronization. On the other hand,
Rappel and Karma \cite{R12b} studied properties of the power spectra of globally coupled neurons 
and found a new effect of noise-induced delta-peak.
Recently, the synchronization 
and  the effect of  CR  in two coupled excitable oscillators are also investigated 
numerically and experimentally \cite{R12e}.

In this paper, the CR of the globally coupled HH neurons
is studied numerically for the first time. We show that the coupling of the network can
enhance CR in two different ways. When
the coupling is weak, the CR phenomenon behaves similar to that of
a single neuron, and no spatiotemporal order can be observed.
When the coupling becomes strong enough, the local measure of coherence jumps up to
a wide plateau first and then jumps down from the plateau as
the intensity of noise increases,
due to the spatiotemporal synchronization of the network.
The coupling tends to
stabilize the noise-induced limit cycle and 
synchronization. The peak frequency of noise-induced limit cycle is
selected to be the spatiotemporal order
through the optimization among the excitability
of a single neuron, the coupling of the network, and the local noise.
The phase of synchronized oscillation is also determined
through the dynamical evolution of the system.
Because the HH model serves as a paradigm
for spiking neurons, we may relate our results to the
existence of coherent spontaneous oscillations
observed in the brain cortex\cite{R13,R14,R15}.

A network of coupled HH neurons is described by the following
equations:

\begin{eqnarray}
\frac{dV_{i}}{dt} &=& f_{i}-I_{i}(t)-\eta_{i}-\frac{1}{N-1}\sum\limits_{j=1,j\neq i}^{N} J_{ij}S_{j},\\
\frac{dm_{i}}{dt} &=&\frac{m_{\infty }(V)-m_{i}}{\protect\tau _{m}(V)}, \\
\frac{dn_{i}}{dt} &=&\frac{n_{\infty }(V)-n_{i}}{\protect\tau _{n}(V)}, \\
\frac{dh_{i}}{dt} &=&\frac{h_{\infty }(V)-h_{i}}{\protect\tau _{h}(V)},
\end{eqnarray}
where $f_{i}=f_{i}(V_{i},m_{i},n_{i},h_{i})$ is
\begin{eqnarray}
f_{i}&=& -g_{Na}m_{i}^{3}h_{i}(V_{i}-V_{Na})
-g_{K}n_{i}^{4}(V_{i}-V_{K})-g_{L}(V_{i}-V_{L}).
\end{eqnarray}
Each neuron is described by a set of four time-dependent variables ($V_{i}$,
$m_{i}$, $n_{i}$, $h_{i}$) where $V_{i}$ is the membrane potential,
$m_{i}$ and $h_{i}$ the activation and inactivation
variables of sodium current, and $n_{i}$ the activation
variable of potassium current. The meaning and detailed
values of the parameters can be found in Ref. \cite{R16}.
The simulation was done by using the fourth order Runge-Kutta
method with the time step being taken as 0.01msec.

Each neuron is subject to an independent noise $\eta_{i}$ with
the same intensity, which is determined from an Ornstein-Uhlenbeck
process $\tau _{c}{d \eta _{i}}/{dt} =-\eta _{i}+\sqrt{2D}\xi$,
where $\xi$ is the Gaussian white noise \cite{R17}. $D$
and $\tau _{c}$ ($=0.1   msec.$) are the intensity and the
correlation time of the noise, respectively. $I_{i}(t)$ is
the input current, which will be time-independent
and will bias the neuron near the saddle-node
bifurcation. The last term in Eq. (1) is the coupling of the network.
The effect of the firing activity of $j$th neuron on the $i$th neuron
is modeled by an impulse current to the $i$th neuron, which is proportional
to the efficacy of the synapse $J_{ij}$ and is generated when the $j$th
neuron is active. $J_{ij}=J$ for all pairs of neurons  with $J$ 
the coupling strength of the system.
The neuron is active whenever its
membrane potential exceeds a threshold $V^{*} $ ($=0 mV$ here).
This activity can be denoted by $S_{j}=\Theta (V_{j}-V^{*})$,
where $\Theta (x)=1$ if $x \geq 0$ and $\Theta (x)=0$ if $x<0$. In the
present simulation, only the excitatory coupling is considered ($J>0$),
that is, the last term is the excitatory postsynaptic potential (EPSP)
received by the single neuron.

The HH neuron is an excitable one. For a dc input
current $I_{0}$, the firing threshold is $I_{c}=6.2 \mu A/cm^{2}$.
The spike limit cycle occurs at $I_{c}$ due to the saddle-node bifurcation.
To observe the CR, we set the input current
$I_{0}=6.0 \mu A/cm^{2}$ for each neuron \cite{R18}, that is,
the system is set in the subthreshold regime near the
threshold or saddle-node bifurcation. For one single HH neuron,
the coherence resonance was discussed in detail in Ref.\cite{R5}.
In the present simulation, we focus on a globally coupled network, and
attempt to extract more significant information of CR.

 The CR exhibits  two different behaviors when
the coupling intensity changes.
They can be seen in the power spectrum of the output spike trains.
In the absence of noise, a single neuron stays at the quiescent state
in which the membrane potential is below $V^{*}$. In this case, there would
be no synaptic transmission between the neurons, and the whole network would
stay at the quiescent state. If an independent local noise ($D\geq 0.3$)
is applied to each neuron, the system begins to fire spike trains.
When the coupling is weak ($\sl e. g. J=5.0$), the power spectrum
densities of the spike trains for different intensities of noise are shown
in Fig. 1(a).  A broad peak can be seen,
similar to the single neuron case (see Fig. 2 in Ref.\cite{R5}).
This behavior of CR is similar but different to that of a single neuron.
\\
\\
\begin{figure}[btp]
\epsfxsize=8.5cm
\epsfbox{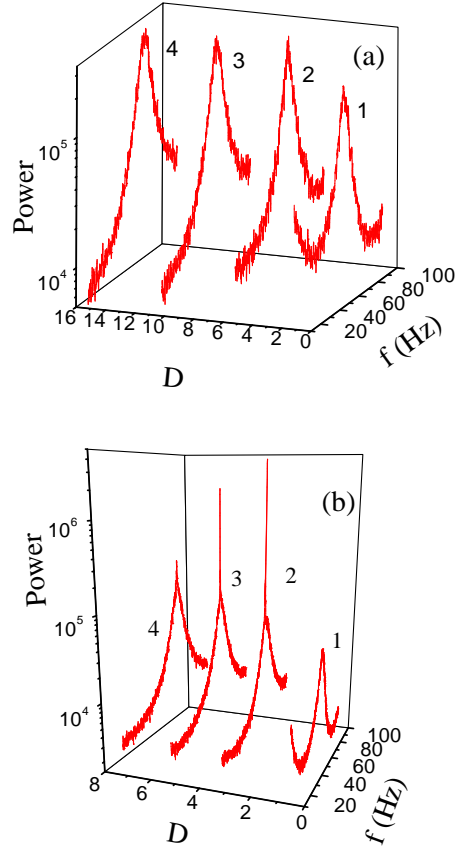}
\caption{(a) The power spectrum of the spike trains with a weak
coupling strength $ J=5.0$  for the noise intensity
$D=1.0$, $5.0$, $10.0$ and $15.0$.
(b) The power spectrum of the spike train with a strong coupling $ J=10.0$
for$ D=0.5$, $3.0$, $5.0$, and $7.0$. The size of the network $N=1000$.}

\end{figure}

When the coupling of the network is strong ($\sl e. g., J=10.0$),
the power spectrum densities of the spike trains for different
intensities of noise are shown in Fig. 1(b). As the noise is weak,
a broad peak is also observed. However, when the noise intensity
increases, the peak becomes higher and sharper. This type of
power spectrum is quite different from that for usual CR
discussed previously. 
The sharp peak is induced by the network itself
and locked at the frequency of spontaneous limit cycle.
The detail of this kind of power spectrum has been addressed in Ref.\cite{R12b}.
When the noise intensity increases further,
the sharp peak tends to become broad, keeping
the general trend of CR in the single neuron case.

The difference of spatiotemporal orders of the network
leads to such two different behaviors of CR. In previous
studies of the conventional SR ,
each unit in the network receives a common external
signal with the same frequency and phase. The external signal
represents an external `clock' leading to the synchronization of the whole
system. So the tuning of the synchronization to the local
noise, which coincides with the local SNR behavior, can  be
observed when the external signal is sufficient strong\cite{R8a}. However, 
in the case of CR, the situation is
different. There is no such kind of global tuning in the
network. The local oscillation of each unit is noise-induced limit cycle.
The phase is random in time and is irrelevant to each other.
Besides,  a broad peak in Fig.1(a) means that the frequency
has some uncertainty. As a result, the synchronization
is not guaranteed in the case of CR.

\begin{figure}[btp]
\epsfxsize=8.5cm
\epsfbox{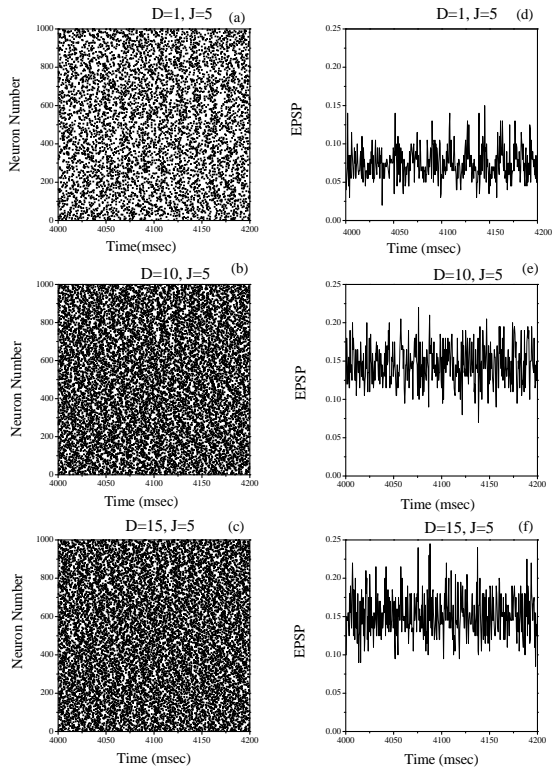}
\caption{
The raster of the network and corresponding excitatory postsynaptic
potential (EPSP) of a neuron with J=5.0 for different intensities of noise:
{D=1.0 ((a) \& (d)), D=10.0 ((b) \& (e)), and D=15.0 ((c) \& (f))}. The network
size N=1000.
}
\end{figure}

When the coupling is weak, the raster records all
the firing events in the network  and the corresponding
EPSP of a single neuron for different intensities of
noise are shown in Fig. 2. From Figs. 2(a)-(c), we
can see that there is no synchronization in the system.
Especially, Fig. 2(b) appears to be the most
coherent state (D=10.0, shown in Fig. 4(a) later).
To see the influence of the network on the local unit, the EPSP
of an arbitrarily chosen neuron is shown in Figs. (d)-(f).
There is a tendency that the EPSP increases when the intensity of
noise increases. The power spectrum of the EPSP  has a broad peak, which coincides with
the CR frequency,  similar to that of the spike train (not shown here).

Figure 3 illustrates how the synchronization can be observed
when the coupling is strong. It is shown in
the raster (Figs. 3(a)-(c)) that,
when the noise is weak (D=0.5), there is no synchronization. Its
corresponding power spectrum is given in line 1 in Fig. 1(b).
When the noise intensity increases, as shown in Fig. 3(b),
the synchronization can be observed. {\it Note that this
spatiotemporal order is achieved by increasing the intensity
of the independent local noise in the absence of
external periodic forcing.} As shown in Fig. 3(e), the EPSP
received by a single neuron has an explicit periodicity,
that is, the network produces a kind of periodic
oscillation due to the synchronization, which is quite similar to a
deterministic signal input to each neuron.  
The corresponding  power spectrum density
of the spike train is shown as line 2 in Fig. 1(b). 
The sharp peak  comes from the periodic EPSP which reflects the effect of the
synchronization on the local unit, in agreement with the work on the coupled integrate 
and fire neurons \cite{R12b}. 
When the noise intensity
increases further, the synchronization is
destroyed; both the explicit periodicity
of the EPSP and the high peak in the power
spectrum of the spike train disappear.

Physically, the spatiotemporal order is established through the
dynamical evolution of the system.
As shown in Eq. (1), the EPSP that each neuron
receives is the average of the events of the other $N-1$ neurons.
Even if there is no synchronization in the system,
the power spectrum of the resulted EPSP should have a dominate frequency
of the limit cycle. This noise-induced EPSP is 
aperiodic. Its intensity and
quality are dependent on the intensity of noise and the coupling strength.
When the coupling strength is weak, the EPSP is very small in comparison
with the intensity of the local noise. No correlation between the
output spike train and the input EPSP can be  established.
When the coupling strength is strong enough, the situation will be different.
Although the EPSP is still too small for a weak noise,
 the quality of EPSP is improved and the intensity
is increased as the noise increases, due to the CR in the
single element level.
Since the input current contains a signal with the same frequency as
the output, the output as well as the EPSP will be refined.
This is a process of positive feedback.
Because the EPSP is the average output
of other neurons, the local neuron tends to keep the pace of such an
averaged signal through the dynamical optimization process.
Finally, a spatiotemporal order can be reached and the frequency of
oscillation, which is just the frequency of CR, is `selected'
by the dynamical process. If the noise intensity increases further,
the synchronization is destroyed.
So the EPSP can be viewed as a kind of indirect feedback.
The EPSP is noise-induced and can be optimized by noise,
while such local noise disturbs the feedback by adding
irregularity at each time step. On the other hand,
when the coupling is significant,
the positive feedback is established. As a result, the EPSP
will evolve gradually to become an identical periodic forcing on every
single element in the system. The synchronization can be observed and
optimized by the noise.  Due to the feature of CR in the globally coupled 
neurons, regardless of whether the system is in the synchronized or desynchronized state, 
the frequency locking at the CR frequency always exists. The synchronization
shown in Fig. 3(b) is a kind of phase locking of all the elements in the network.

\begin{figure}[btp]
\epsfxsize=8.5cm
\epsfbox{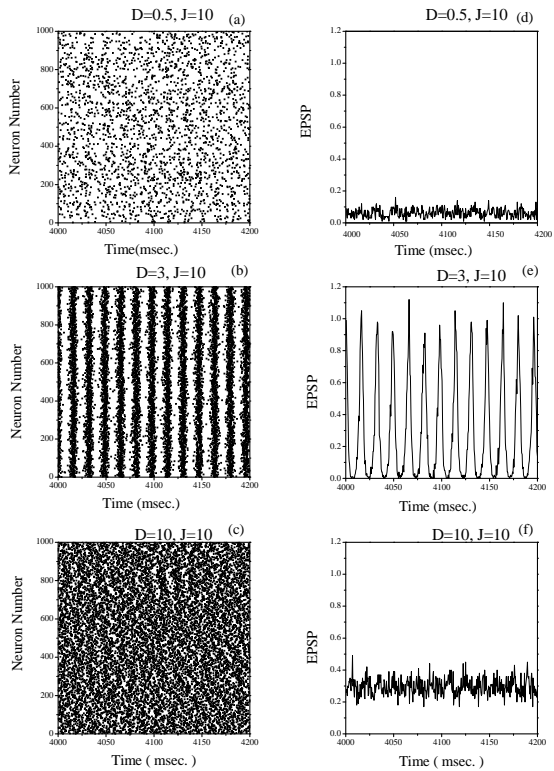}
\caption{
The raster of the network and corresponding excitatory postsynaptic
potential (EPSP) of a neuron with J=10.0 for different intensities of noise:
{D=0.5 ((a) \& (d)), D=3.0 ((b) \& (e)), and D=10.0 ((c) \& (f))}. The network
size N=1000.
}
\end{figure}

Such noise-induced synchronization possesses two interesting features. First, the
synchronization frequency is dependent on the local noise and the coupling.
Secondly, the phase of spatiotemporal oscillation is determined by the
dynamical evolution of the system itself.
Because of this, 
the peak frequency of CR is locked at the frequency of
the synchronized oscillation. However, the phase of the synchronized
oscillation is `selected' by the indirect feedback process which
is sensitive to the detail process in the noisy environment. For example,
different initial conditions of the simulation lead to the same frequency
but different phases of the synchronized oscillation.

We can characterize CR quantitatively via a coherence factor
$\beta $\cite{R2}, which is the measure of coherence and defined as:

\begin{eqnarray}
\beta =h(\Delta \omega /\omega _{p})^{-1},
\end{eqnarray}
where $h$ and $\omega _{p}$ are the height  and the frequency of the peak,
and $\Delta \omega$ is
the width of the peak at the height $h_{1}=e^{-\frac{1}{2}} h$.

The $\beta$ $\sl vs$ the noise intensity $D$ for different couplings of the
network is shown in Fig. 4(a).
When $D$ increases, $\beta$ increases first and then decreases
after reaching the maximum. The coupling may be viewed
as a tuning parameter of CR. For comparison, the CR of a single neuron
case is also displayed in the figure (J=0). The
enhancement of CR is significant when the coupling is stronger.
When the coupling is weak, there is no spatiotemporal order in the system.
The value of $\beta$ is the same order of the magnitude as that of the
single neuron case, and similar $\beta -D$ curves are exhibited in the two
cases. However, when the coupling becomes strong enough,
the $\beta$ increases dramatically with D at first, showing the onset of
synchronization, and then a wide plateau is followed,
indicating that the self-evolved
spatiotemporal order is stable against a large range intensity of local noise.
The normalized $\beta$ $\sl vs$ the noise intensity for different coupling
is also shown in the inset of Fig. 4(a) .

The difference of the CR in the single neuron case and the coupled neurons can
be seen in Fig. 4(b), in which the peak height of the power spectrum densities of the
spike train is plotted against the noise intensity $D$ for different couplings of the 
network. In the single HH neuron case (J=0), the height of the peak 
increase monotonically as the noise increases 
(see also Figure 4(b) in Ref.\cite{R5} ). 
In the coupled HH neurons, similar to 
Fig. 4(a), a bell-shaped curve is observed. Once the synchronization is established, the peak 
height increases dramatically. On the other hand, even when the coupling is weak and no synchronization 
is established, as shown in the inset  of Fig. 4(b), the bell-shape curve can still be 
observed (J=1, and J=5 curve in Fig. 4(b)).  This means that the height of CR peak is 
tuned by  the noise  in the absence  of  synchronization. 
As shown in Fig. 2(d)-(f), the EPSP can 
be regarded as a kind of aperiodic signal which has the same frequency
as the output. The tuning to the noise of such an aperiodic signal is similar to SR, however,
unlike the usual SR, the EPSP here is produced by the network itself through CR. The 
intensity and quality of the EPSP are different for different strengthens of noise due to the effect of CR.  
So, even though the power spectrum  density of the spike train is similar to that of the 
single neuron case,  the mechism is different. The effect of CR can be enhanced significantly
by the coupling  even when  there is no synchronization.

\begin{figure}[btp]
\epsfxsize=8.5cm
\epsfbox{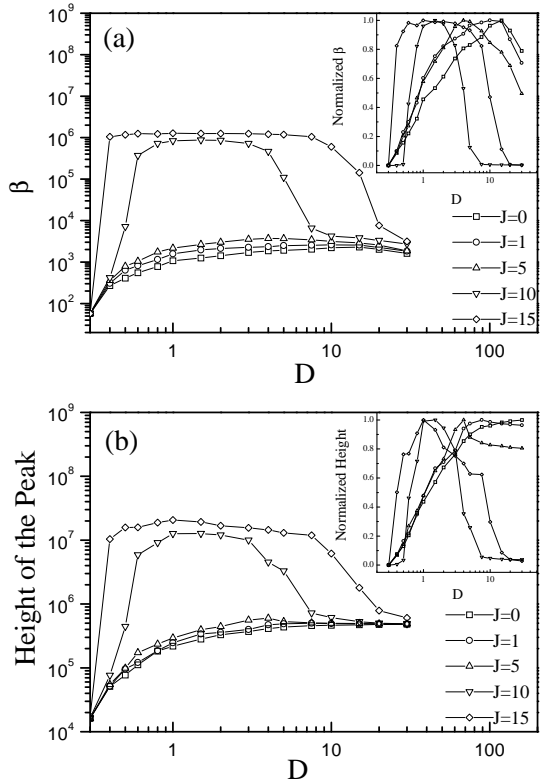}
\caption{
(a) The measure of coherence  $\beta$ versus the intensity of noise for different
coupling strengths. Inset: The normalized
coherence factor $\beta$ versus the intensity of noise. The same data in (a) is
divided by its own maximum for each curve. 
(b) The height of the peak of  the power spectrum versus the intensity of
noise for different coupling strength. Inset: The normalized height
of peak versus the intensity of noise. The same data in (b) is divided by its own
maximum for each curve. The size of the network is N=100.
The lowest lines in (a) and (b) are the same one for the single neuron case.
}
\end{figure}

Figure 5(a) illustrates how the $\beta$ changes with the size of
the strongly coupled network (J=10.0).
Clearly, the $\beta -D$ curve changes little whenever
the number of the neurons in the network is larger than 50,
with the onset-point and the end-point of synchronization being
almost unchanged. Although the network is
globally coupled, the degree of synchronization is roughly
irrelevant to the size of the network if it is sufficiently large.

Figure 5(b) shows the peak frequency of CR as a function of the
intensity of noise for different coupling strengths.
We can see that, regardless of the coupling strength,
the frequency will increase when the noise increases, with the same tendency
as that for a single neuron case. On the other hand, the frequency
increases as the coupling strength increases, 
tuning CR in another way. Moreover, There is no dramatic change of
the frequency when the spatiotemporal order is established. In fact,
we can not see the difference of synchronized and non-synchronized
states of the system from this kind of plot. Both are CR states.

\begin{figure}[btp]
\epsfxsize=8.5cm
\epsfbox{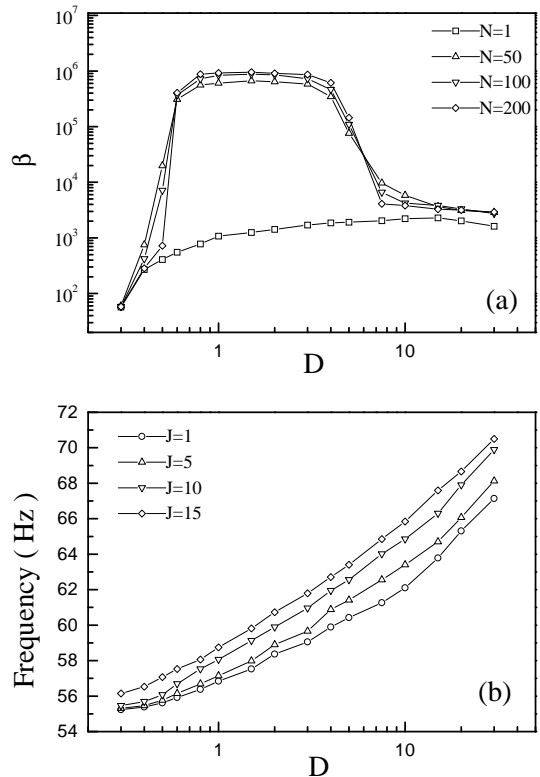}
\caption{
(a) The measure of coherence $\beta$ versus the intensity of noise for different sizes of the 
network when J=10.0.
(b) The frequency of CR versus the noise intensity for different coupling
strengths. The size of network N=100.
}
\end{figure}

Finally, we address the relevance of the CR of the globally
coupled HH neurons to the activities of realistic
neural systems. In recent years, synchronized
spontaneous oscillations have been observed in the brain cortex and are
proposed to possess a binding function, where the spatially-distributed
neurons resonate to generate large function states that bring about
cognition \cite{R13,R14,R15}. From the simulations, we may elucidate
how these synchronized spontaneous oscillations are established. It would
be the CR state. The frequency of oscillation is determined by the
excitability of a single neuron, the coupling of the network, and
the noise. On the other hand, the synchronization may be noise-induced,
giving a possibility that the noise would play an active role in neural
activities.  The synchronized state would be stable in
a large range intensity of the local noise. This feature
would enable the neural system to fulfill cognition function
in noisy environment.

In summary, we have studied the CR of globally coupled network of
HH neurons. It is found that, when the coupling is strong,
the synchronization is induced and optimized by the noise.
The frequency of CR of the local element is locked at  the
spatiotemporal oscillation frequency, and the phase of spatiotemporal oscillation
is determined by  the dynamical evolution. A wide
plateau in the $\beta -D$ curve was observed for the strongly
coupled network with large sizes, indicating a stable
spatiotemporal order in a large range intensity of local noise. 
The effect of CR can be enhanced greatly by the coupling regardless of the 
spatiotemporal order of the system.
Our results may be relevant to
the synchronized spontaneous oscillations observed in some realistic
neural systems.


\begin{references}
\bibitem{R18} It is not necessary to assume the same $I_{0}$ for every
              neuron in the network.
             The only requirement is to set the neurons near the threshold
             or saddle-node bifurcation. For a random
             distributed $I_{0}$, we will observe almost the same
             phenomenon reported later on.
\end{references}
\end{document}